\documentclass[preprint,showpacs,preprintnumbers,amsmath,amssymb]{revtex4}
\usepackage{booktabs}
\usepackage{mathrsfs}
\usepackage{epsfig}
\usepackage{graphicx}
\usepackage{dcolumn}
\usepackage{bm}
\usepackage{amsmath}

\let\jnfont=\rm
\def\NPB#1,{{\jnfont Nucl.\ Phys.\ B }{\bf #1},}
\def\PLB#1,{{\jnfont Phys.\ Lett.\ B }{\bf #1},}
\def\EPJC#1,{{\jnfont Eur.\ Phys.\ J.\ C }{\bf #1},}
\def\PRD#1,{{\jnfont Phys.\ Rev.\ D }{\bf #1},}
\def\PRL#1,{{\jnfont Phys.\ Rev.\ Lett.\ }{\bf #1},}
\def\MPLA#1,{{\jnfont Mod.\ Phys.\ Lett.\ A }{\bf #1},}
\def\JPG#1,{{\jnfont J.\ Phys.\ G}{\bf #1},}
\def\CTP#1,{{\jnfont Commun.\ Theor.\ Phys.\ }{\bf #1},}
\def\ZPC#1,{{\jnfont Z.\ Phys.\ C }{\bf #1},}
\def\JHEP#1,{{\jnfont JHEP \ }{\bf #1},}
\def\Rv{\not{\hbox{\kern-1pt $R$}}}
\def\p{\not{\hbox{\kern-3pt $p$}}}
\begin{document}

\title{Top quark three-body decays in $R$-violating MSSM}

\author{ Zhaoxia Heng$^1$, Gongru Lu$^2$, Lei Wu$^2$, Jin Min Yang$^1$ }

\affiliation{
$^1$ Key Laboratory of Frontiers in Theoretical Physics, Institute of Theoretical Physics,
     Academia Sinica, Beijing 100190, China \\
$^2$ College of Physics and Information Engineering, Henan Normal University, Xinxiang 453007, China
     \vspace*{1.5cm}}

\begin{abstract}
In the minimal supersymmetric standard model the R-parity
violating interactions can trigger various exotic three-body
decays for the top quark, which may be accessible at the LHC. In
this work we examine the R-violating decays  $t \to c X_1 X_2$,
which include the tree-level processes $t \to c \ell^-_i \ell^+_j$
($\ell_i=e, \mu, \tau$) and $t \to c d_i \bar{d}_j$ ($d_i=d, s,
b$), as well as the loop-induced processes $t \to c g X$ ($X=g,
\gamma, Z, h$). We find that the hereto weakly constrained
R-violating couplings can render the decay branching ratios quite
sizable, some of which already reach the sensitivity of the
Tevatron collider and can be explored at the LHC with better
sensitivity.

\end{abstract}
\maketitle

\section{INTRODUCTION}
Top quark is one of the forefront topics in high energy physics.
As the heaviest known elementary particle, top quark is speculated
to be a window into the TeV-scale physics.
The properties of the top quark have been being measured at the Tevatron
collider
and so far the Tevatron data are in agreement with the SM predictions.
However, due to the small statistics of the Tevatron collider, the precision
of current measurements of the top quark properties is not so good and hence
there remains plenty of room for new physics in the top quark sector.
The CERN Large Hadron Collider (LHC) will serve as a top quark factory and
allow to scrutinize the top quark nature. The precise measurement of the top quark
properties at the LHC may provide clues to new physics beyond the Standard Model (SM)
\cite{topreview}.

Due to the large number of top pair samples at the LHC, various exotic decays of
the top quark could be explored with high sensitivity. In the SM top quark dominantly
decays into a $W$-boson plus a bottom quark. But in new physics models various
exotic decays can open up and may reach the detectable level.
For example, the FCNC
two-body decays $t\to c X$ ($X=g,\gamma,Z,h$), which are extremely small in the
SM \cite{tcvh-sm},
could be enhanced to the observable level in the minimal supersymmetry model
(MSSM) \cite{tcv-mssm-1,tcv-mssm-2}
and the technicolor models \cite{tcv-tc2}.
Also, some new decay modes may open up in new physics models,
such as $t\to \tilde\chi^0_1 \tilde{t}$ in the MSSM \cite{hosch}.
Given the stringent lower bounds on the masses of new particles (like the
top-squark $\tilde{t}$ and the lightest neutralino $\tilde\chi^0_1$) from
current experiments, such two-body new decay modes are getting kinematically
suppressed or forbidden. In this work we focus on the kinematically allowed three-body
decays with all the final states being the SM particles.

It is well known that in the MSSM the $R$-parity,
defined by $R=(-1)^{2S+3B+L}$ with spin $S$, baryon-number $B$ and
lepton-number $L$,  is often imposed on the Lagrangian to maintain
the separate conservation of $B$ and $L$.
But this conservation requirement is not dictated by any fundamental
principle such as gauge invariance and renormalizability.
Therefore, the phenomenology of $R$-parity violation has attracted much attention.
For the effects of R-violating interactions in the top quark sector,
we may have large FCNC two-body decays $t\to c X$ ($X=g,\gamma,Z,h$)
and some exotic top productions at the LHC \cite{rv-t-prod}.
Note that the R-violating couplings can also induce various three-body
decays for the top quark.
In this work we examine the three-body decays with all the final states being the SM
particles, which include
the tree-level processes $t \to c \ell^-_i \ell^+_j$ ($\ell_i=e, \mu, \tau$)
and $t \to c d_i \bar{d}_j$ ($d_i=d, s, b$),
as well as the loop-induced processes $t \to c g X$ ($X=g, \gamma, Z, h$).
Although these three-body decays may be quite rare, they are still worth checking
because the top decays can be soon scrutinized at the LHC.
As will be shown by our study,
the hereto weakly constrained R-violating couplings can make
the decay branching ratios quite sizable, some of which already
marginally reach the sensitivity of the Tevatron collider.

Note that the top quark three-body decays with one or two new (heavy) particles
in the final states have been studied in the MSSM with or without R-parity
\cite{t-3decay-mssm1,t-3decay-mssm2,rv-t-decay}
and in the general two-Higgs-doublet model \cite{t-3decay-2hdm}.
In \cite{t-3decay-mssm2} the L-violating decay $t \to c \ell^-_i \ell^+_j$
was also studied.  In our study we include it for completeness.

This work is structured as follows. In Sec. II, we recapitulate the R-parity
violating couplings and present the calculations for top three-body decays
at the LHC. In Sec. III, we show some numerical results for the branching ratios
of these decays. In Sec. IV we draw our conclusion.
The analytic expressions from the loop calculations are presented
in the Appendix.

\section{Calculation}
The R-parity violating superpotential of the MSSM is given by \cite{Rpv}
\begin{eqnarray}\label{potential}
\frac{1}{2}\lambda_{ijk}L_iL_jE_k^c +\lambda_{ijk}'L_iQ_jD_k^c
               +\frac{1}{2}\lambda^{\prime\prime}_{ijk}U_{i}^cD_{j}^cD_{k}^c
\end{eqnarray}
where $L_i(Q_i)$ and $E_i^c(U_i^c,D_i^c)$ are respectively the doublet
and singlet lepton (quark) chiral superfields, and $i,j,k$ are generation
indices. The terms of $\lambda$ and $\lambda^{\prime}$ violate
lepton number while the terms of $\lambda^{\prime\prime}$ violate
baryon number. The non-observation of the proton decay
imposes very strong constraints on the product of the $L$-violating
and $B$-violating couplings~\cite{proton}. Thus in our numerical calculation
we will assume that only one type of these interactions
(either $L$- or $B$-violating) exist.
Since only $\lambda^{\prime}$ or $\lambda^{\prime\prime}$ couplings
can induce the three-body top quark decays, we will drop the $\lambda$
couplings in the following.
In terms of the four-component Dirac notation, the Lagrangian of
$\lambda^{\prime}$ and $\lambda^{\prime\prime}$ couplings is given by
\begin{eqnarray}
{\cal L}_{\lambda^{\prime}}&=&-\lambda^{\prime}_{ijk}
\left [\tilde \nu^i_L \bar d^k_R d^j_L
+ \tilde d^j_L \bar d^k_R \nu^i_L
+ \tilde d^{k*}_R \bar \nu^{ic}_R d^j_L
- \tilde l^i_L \bar d^k_R u^j_L
- \tilde u^j_L \bar d^k_R l^i_L
- \tilde d^{k*}_R \bar l^{ic}_R u^j_L\right]+h.c.\\
{\cal L}_{\lambda^{\prime\prime}}&=&-\frac{1}{2}\lambda^{\prime\prime}_{ijk}
\left [\tilde d^{k*}_R \bar u^{i}_R d^{jc}_L+\tilde d^{j*}_R\bar u^i_Rd^{kc}_L
       +\tilde u^{i*}_R\bar d^j_R d^{kc}_L\right ]+h.c.
\end{eqnarray}

\begin{figure}[htb]
\epsfig{file=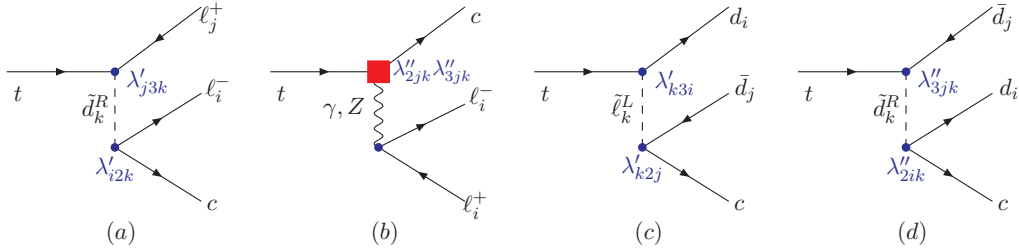,width=14cm}
\vspace*{-0.5cm}
\caption{Feynman diagrams: (a) $t \to c \ell^-_i \ell^+_j$
         induced at tree-level by
         $\lambda^{\prime}_{i2k}\lambda^{\prime}_{j3k}$;
         (b) $t \to c \ell^-_i \ell^+_i$
         induced at loop-level by
         $\lambda^{\prime\prime}_{2jk}\lambda^{\prime\prime}_{3jk}$
         where the effective vertices $tc\gamma$ and $tcZ$
         are similar to the effective vertex $tcg$
         defined in Fig.\ref{fig3};
         (c-d) $t \to c d_i \bar{d}_j$ induced at tree-level by
          $\lambda^{\prime}_{k2j}\lambda^{\prime}_{k3i}$ and
          $\lambda^{\prime\prime}_{2ik}\lambda^{\prime\prime}_{3jk}$,
          respectively. In (a) the charged lepton $\ell^-_i$ and
          the charm quark can be replaced respectively by a neutrino
         $\nu_i$ and a strange quark to give the process
         $t \to s \nu_i \ell^+_j$.}
\label{fig1}
\end{figure}

\begin{figure}[htb]
\epsfig{file=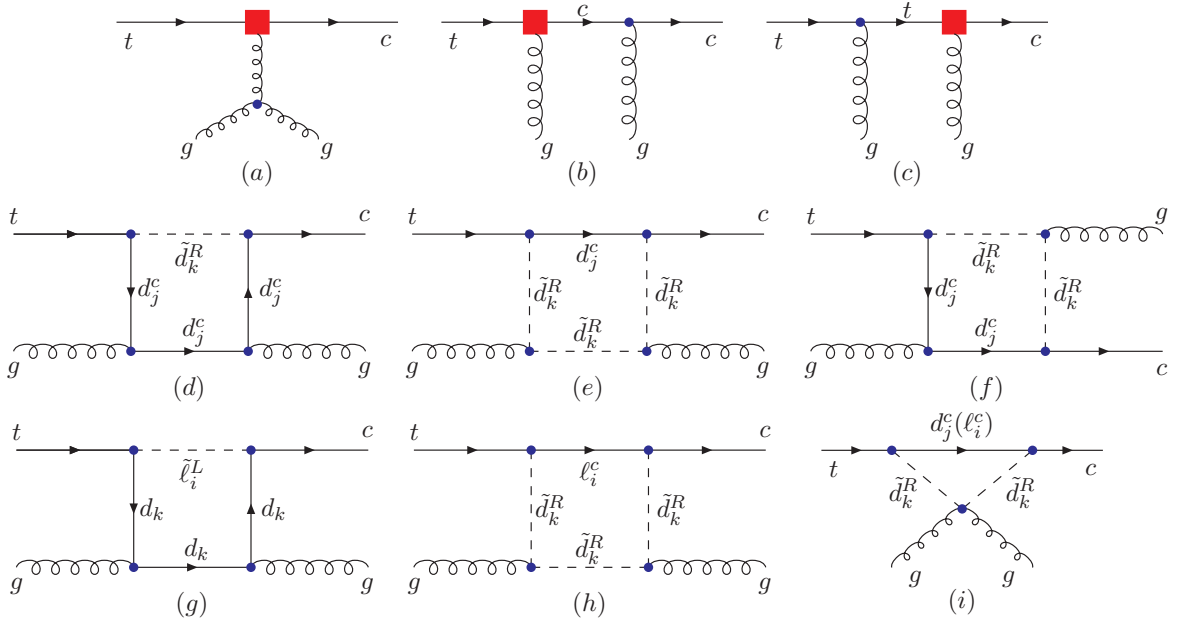,width=16cm}
\caption{Feynman diagrams for the decay
         $t \to c gg$ induced at loop-level by
         $\lambda^{\prime}_{i2k}\lambda^{\prime}_{i3k}$
         or $\lambda^{\prime\prime}_{2jk}\lambda^{\prime\prime}_{3jk}$
         with the effective vertex $tcg$ defined in Fig.\ref{fig3}.}
\label{fig2}
\end{figure}

\begin{figure}[htb]
\epsfig{file=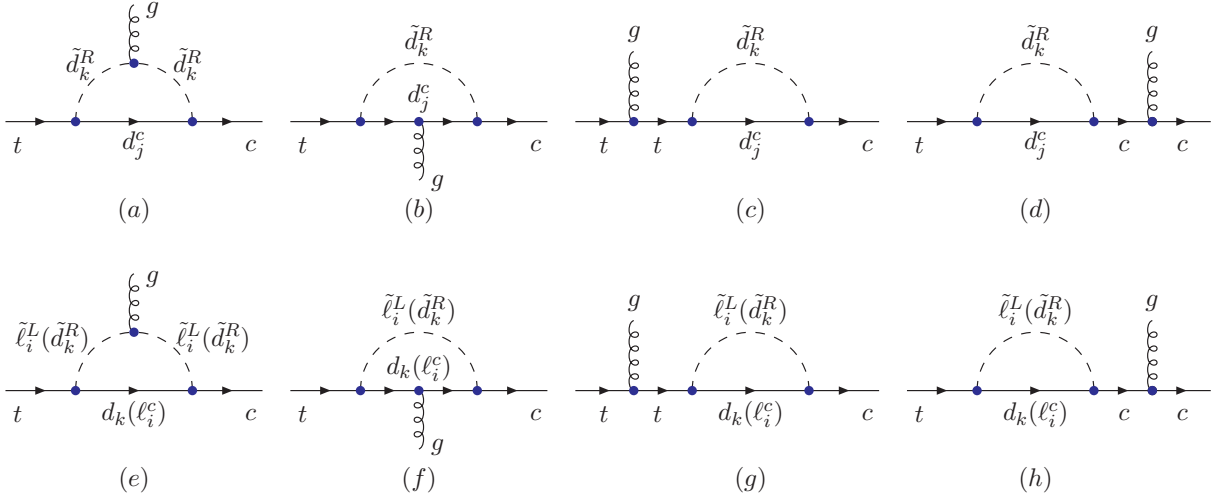,width=16cm}
\vspace*{-0.5cm}
\caption{Feynman diagrams for the effective vertex $tcg$ appeared in Fig.\ref{fig2},
         induced at loop-level by
         $\lambda^{\prime}_{i2k}\lambda^{\prime}_{i3k}$
         or $\lambda^{\prime\prime}_{2jk}\lambda^{\prime\prime}_{3jk}$.}
\label{fig3}
\end{figure}

The three-body decays $t \to c \ell^-_i \ell^+_j$ ($i$ and $j$ can
be equal or not equal) can be induced at tree-level by the
L-violating $\lambda^{\prime}_{i2k}\lambda^{\prime}_{j3k}$, as
shown in Fig.\ref{fig1}(a); while $t \to c \ell^-_i \ell^+_i$ can
also be induce at loop-level by the B-violating
$\lambda^{\prime\prime}_{2jk}\lambda^{\prime\prime}_{3jk}$, as
shown in Fig.\ref{fig1}(b) where the effective vertices $tc\gamma$
and $tcZ$ are similar to the effective vertex $tcg$ defined in
Fig.\ref{fig3}. The decays $t \to c d_i \bar{d}_j$ can be induced
at tree-level either by the L-violating
$\lambda^{\prime}_{k2j}\lambda^{\prime}_{k3i}$ or by the
B-violating
$\lambda^{\prime\prime}_{2ik}\lambda^{\prime\prime}_{3jk}$, as
shown in Fig.\ref{fig1}(c) and (d), respectively. The decay $t \to
c gg$ can be induced at loop-level by the L-violating
$\lambda^{\prime}_{i2k}\lambda^{\prime}_{i3k}$ or the B-violating
$\lambda^{\prime\prime}_{2jk}\lambda^{\prime\prime}_{3jk}$ as
shown in Fig.\ref{fig2} with the effective vertex $tcg$ defined in
Fig.\ref{fig3}. The Feynman diagrams for $t \to cg\gamma, cgZ,
cgh$ are similar to $t \to c gg$ and are not plotted here. The
analytic expressions of the amplitudes for the loop-induced
processes are lengthy and tedious. Here, as an example, we list
the expressions for the L-violating loop contributions to the
effective vertex $tcV$ in Appendix A. The corresponding
B-violating contributions can be found in the third paper in
\cite{tcv-mssm-2}.

The current upper bounds for all R-parity violating couplings are
summarized in \cite{bound}.
Table I is a list of current
limits for those couplings relevant to our study, taken from \cite{bound}.
We see that the constraints are quite weak for the
couplings $\lambda^{\prime\prime}_{2jk}$ and
$\lambda^{\prime\prime}_{3jk}$, which induce the tree-level
decays shown in Fig.1(d).

\begin{table}
\caption{Current upper limits on the R-parity violating couplings
relevant to our study, taken from \cite{bound}.}
 \begin{tabular}{lll} \hline
 couplings & ~~~~~~~~bounds & ~~~~~~~~~~~~~~~~~~~sources \\
 \hline
$\lambda^{\prime\prime}_{212},~\lambda^{\prime\prime}_{213},~\lambda^{\prime\prime}_{223}$
~~~~~~~~~~~~& $~~~~~~~~~1.25 $ &~~~~~~~~~~~~~~~~ perturbativity\\
$\lambda^{\prime\prime}_{312},~\lambda^{\prime\prime}_{313},~\lambda^{\prime\prime}_{323}$
~~~~~~~~~~~~& $ 0.97\times (m_{\tilde d_{kR}}/100{\rm ~GeV})$ &
~~~~~~~~~~~~~~~~~~Z-decays \\
$\lambda^{\prime}_{i1k}$, ~$\lambda^{\prime}_{i2k}$
  ~~~~~& $0.11\times \sqrt{m_{\tilde d_{kR}}/100{\rm ~GeV}}$ &
~~~~~~~~~~~~~~~~$K^0-\bar K^0$ mixing \\
$\lambda^{\prime}_{i3k}$  ~~~~~& $1.1\times \sqrt{m_{\tilde d_{kR}}/100{\rm ~GeV}}$  &
~~~~~~~~~~~~~~~ $B_d-\bar B_d$ mixing \\
$\lambda^{\prime}_{i2k}\lambda^{\prime}_{i3k}$
~~~~~~~~~~~~& $0.09\times (m_{\tilde d_{kR}}/100{\rm ~GeV})^2 $ &
~~~~~~~~~~~~~~~~~~ $B\to K\gamma$\\
\hline
 \end{tabular}\label{dlambda}
 \end{table}

\section{NUMERICAL RESULTS AND DISCUSSIONS}
In our calculation the top quark mass is taken as the new CDF value
$m_{t}=172$ GeV \cite{CDF-mt}
Other SM parameters are taken as \cite{pdg}
$m_{Z}=91.19 {\rm ~GeV}$, $m_{W}=80.4 {\rm ~GeV}$,
$\sin^2\theta_W=0.2228$, $\alpha_s(m_t)=0.1095$ and $\alpha=1/128$ \cite{pdg}.
The SUSY parameters involved in our calculations are the masses of squarks and
sleptons as well as the R-parity violating couplings $\lambda^{\prime}_{i2k}$,
$\lambda^{\prime}_{i3k}$,$\lambda^{\prime\prime}_{2jk}$ and
$\lambda^{\prime\prime}_{3jk}$, whose upper bounds are listed in
Table I. The strongest bound on squark mass
is from the Tevatron experiment. For example, from the search for the
inclusive production of squark and gluino in R-conserving minimal supergravity
model with $A_0 = 0$, $\mu< 0$ and $\tan \beta = 5$, the CDF gives a bound
of 392 GeV at the 95 $\%$ C.L. \cite{CDF} for degenerate gluinos and squarks.
However, this bound may be not applicable to the R-violating scenario because
the SUSY signal in case of R-violation is very different from the R-conserving
case. The most robust bounds on sparticle masses come from the LEP results, which
give a bound of about 100 GeV on squark or slepton mass \cite{LEP}.
In our numerical calculations we assume the presence
of the minimal number of R-violating couplings, i.e.,
for each process only the two relevant couplings (not summed
over the family indices) are assumed to be present.

\begin{figure}[htb]
\epsfig{file=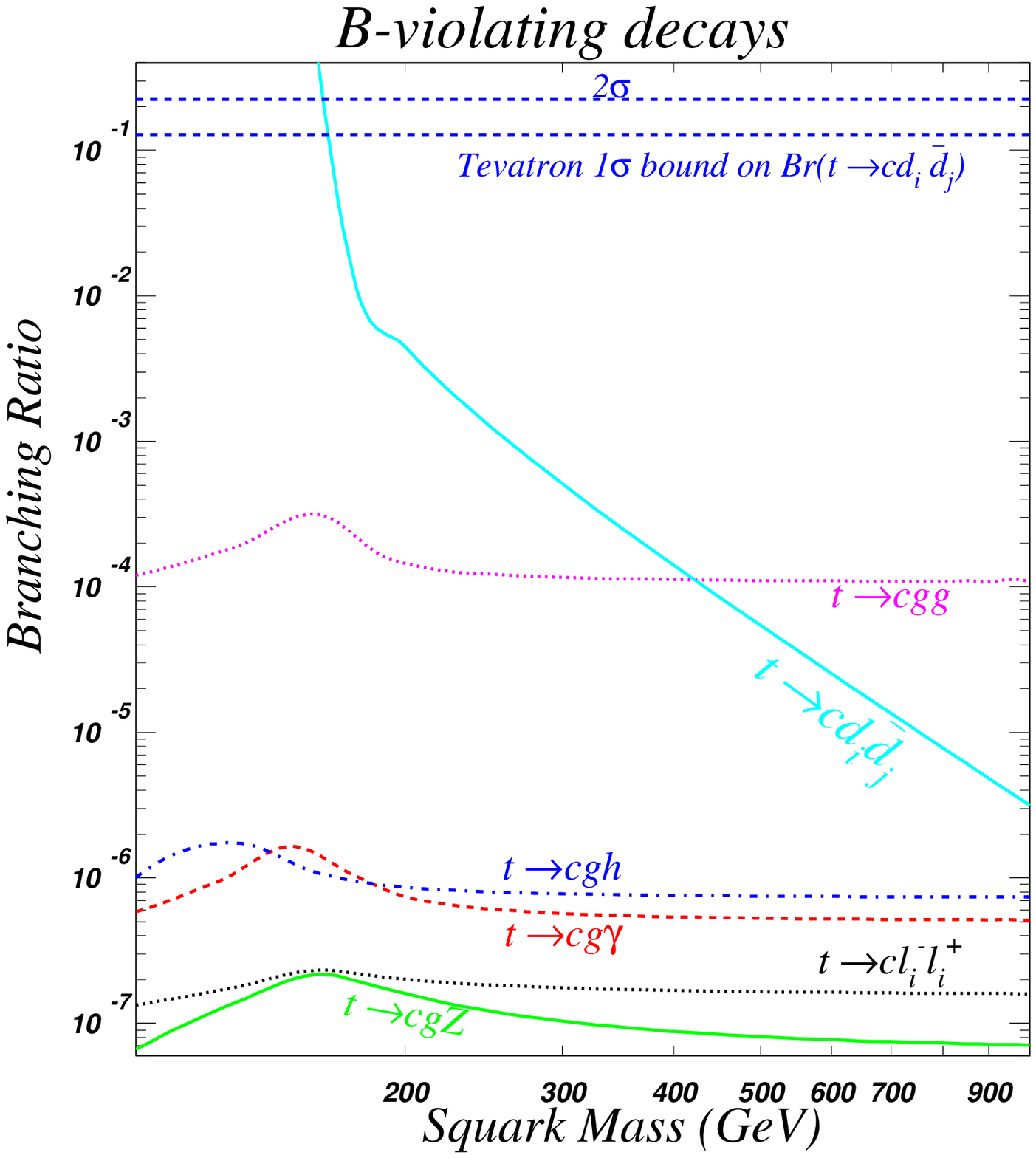,width=8.3cm}
\hspace{-0.5cm}
\epsfig{file=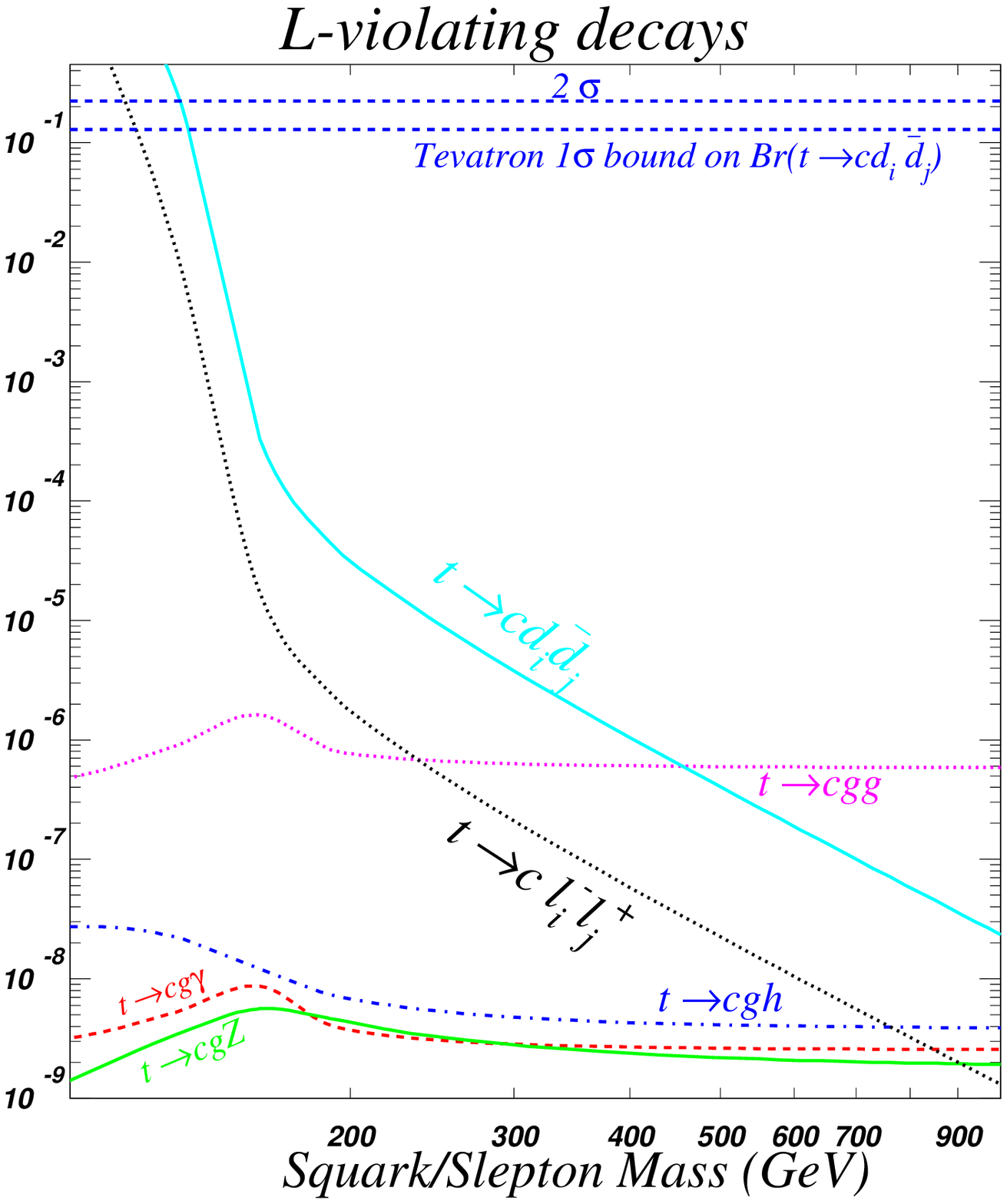,width=8.3cm}
\vspace*{-0.5cm}
\caption{The branching ratios of the R-violating three-body decays of the top quark
         as a function of squark or slepton mass. In the left frame the
         product of the two $\lambda^{\prime\prime}$ couplings involved in
         each decay is taken as 1.2, while in the right frame the
         product of the two $\lambda^{\prime}$ couplings involved in
         each decay is taken as 0.1.
         In the right frame the curve of $t \to c \ell^-_i \ell^+_j$
         also applies to  $t \to s \nu_i \ell^+_j$. }
\label{fig4}
\end{figure}

In Fig. \ref{fig4} we show the branching ratios of the three-body decays
as a function of squark or slepton mass.
In this figure the product of the two $\lambda^{\prime\prime}$ ($\lambda^{\prime}$)
couplings involved in each decay is fixed as 1.2 (0.1), which
are approximately the maximal values shown in
Table I for squark or slepton mass of 100 GeV.

Among the B-violating decay modes shown in the left frame of Fig. \ref{fig4},
$t\to c d_i \bar d_j$ has the largest
branching ratio because it is a tree-level process, as shown in Fig1.(d);
while other decay modes are all induced at loop-level,
among which $t\to c gg$ has the largest branching ratio.

Among the L-violating decay modes shown in the right frame of Fig. \ref{fig4},
$t\to c d_i \bar d_j$ also has the largest
branching ratio because it is a tree-level process, as shown in Fig1.(c).
The decay $t\to c \ell^-_i \ell^+_j$ also occurs at tree-level, as shown in Fig1.(a);
but its branching ratio is always below $t\to c d_i \bar d_j$ for a common value
of slepton and squark mass because  $t\to c d_i \bar d_j$ is relatively enhanced
by a color factor. Other decay modes are all induced at loop-level, among which
$t\to c gg$ has the largest branching ratio.

We also calculated the channels $t \to c \gamma X (X=\gamma,Z,h)$
and found that their branching  ratios are below $10^{-9}$,
which are far below the detectable level of the LHC and thus
are not plotted in the figures.

\begin{figure}[htb]
\epsfig{file=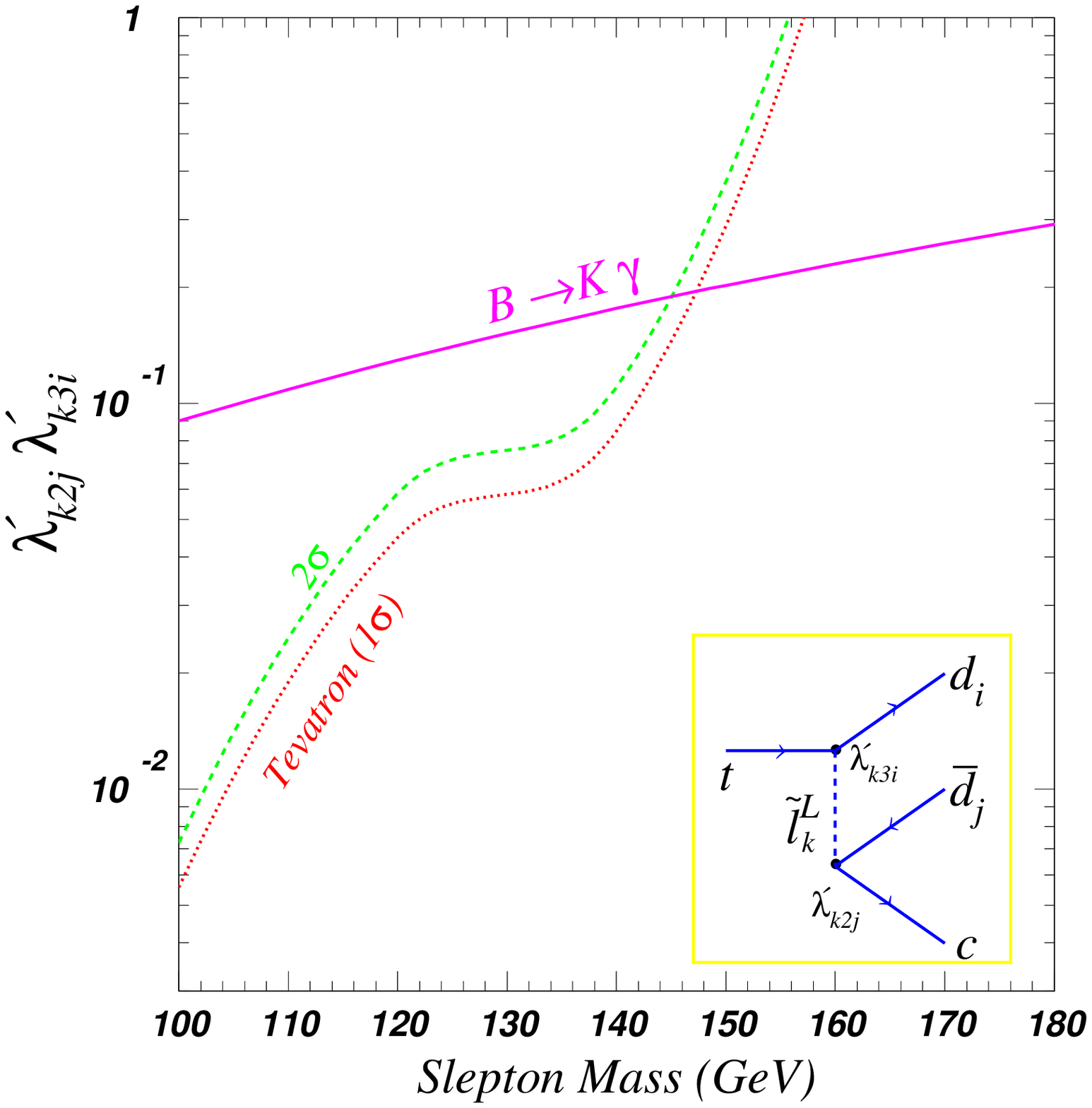,width=8.1cm}
\epsfig{file=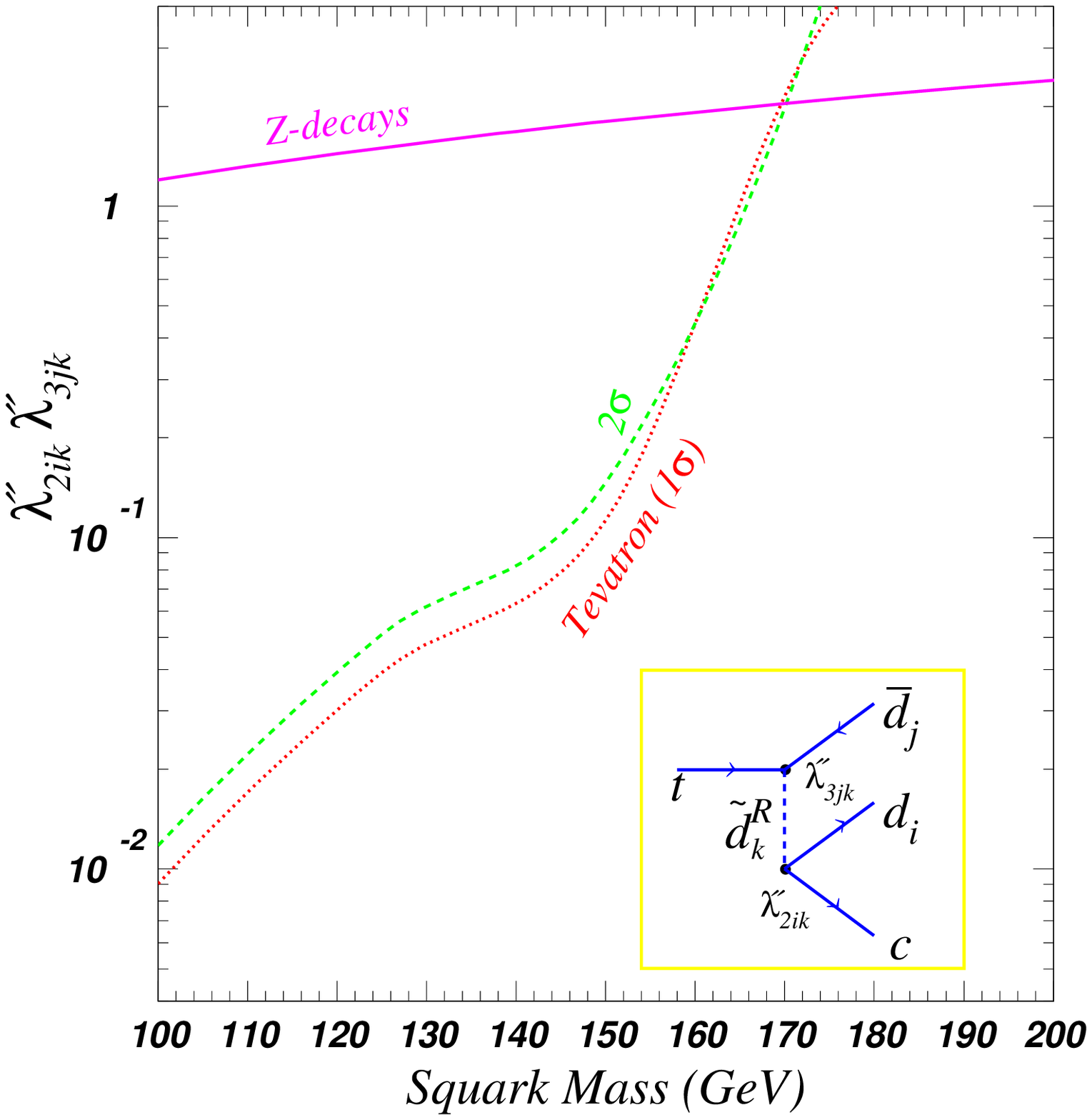,width=8.1cm}
\vspace*{-0.5cm}
\caption{The Tevatron bound shown on the plane of the relevant $\lambda^{\prime\prime}$
versus squark mass or $\lambda^{\prime}$ versus slepton mass. The bounds from
$B\to K \gamma$ and $Z$-decays \cite{bound} are also plotted for comparison.
Above each
curve is the corresponding excluded region.}
\label{fig5}
\end{figure}

From  Fig. \ref{fig4} we see that for a squark or slepton mass
below 200 GeV, the decay $t\to c d_i \bar d_j$ can be quite sizable
and could even compete with the SM decay $t\to W^+ b$. Such a large
branching ratio  could be readily constrained
by the available data at the Tevatron.
For example, the decay $t\to c d_i \bar d_j$ is 'exotic' to the
dileptonic $t\bar{t}$ event counting because the final states
of $t\bar t$ followed by $t\to c d_i \bar d_j$ and/or
$\bar{t}\to \bar{c} \bar{d}_i d_j$ do not have enough leptons
to be included in the dileptonic event samples.
In other words, only the normal decay modes of $t$ and $\bar{t}$
(i.e., $t\to W^+ b$ and $\bar{t}\to W^- \bar{b}$) can be counted
into the dileptonic event samples.
By comparing the CDF data \cite{CDF-tt}
$\sigma[t\bar t]_{\rm exp}=6.7\pm 0.8(stat)\pm 0.4(syst)\pm 0.4(lumi)$ pb
measured from dileptonic channels with
$\sigma[t\bar t]_{\rm QCD}[1-Br(t\to c d_i \bar d_j)]^2$,
we find that the upper bound on
$B(t\to c d_i \bar d_j)$ given by
\begin{equation}
Br(t\to c d_i \bar d_j)
\le \left \{ \begin{array}{ll} 0.13~ & ~~(1\sigma) \\
                               0.22~ & ~~(2\sigma)
               \end{array} \right.
\end{equation}
where we used $\sigma[t\bar t]_{\rm QCD}=7.39^{+0.57}_{-0.52}$ pb \cite{QCD-tt}
and neglected the SUSY effects on the production rate \cite{tt-susy}.
Such an upper bound is plotted as the horizontal lines in Fig. \ref{fig4}.
If we project the bound on the plane of the
$\lambda^{\prime\prime}_{2ik}\lambda^{\prime\prime}_{3jk}$
versus squark mass or $\lambda^{\prime}_{k2j}\lambda^{\prime\prime}_{k3i}$
versus slepton mass, we obtain
Fig. \ref{fig5}, where we also show the bounds from $B\to K \gamma$ and
$Z$-decays \cite{bound}. We see that the Tevatron has a better sensitivity for
a light squark or slepton.

Since the statistical uncertainty of the top production rate will be
greatly reduced at the LHC, the LHC will have better sensitivities
to these exotic three-body decays.
Of course, the sensitivity will be different for different decay channels.
To figure out the sensitivity for each channel we need detector-dependent
full Monte Carlo simulations. A preliminary fast detector simulation showed
\cite{rv-t-decay-lhc}
that the LHC may have a high sensitivity (about $10^{-6}$) to the R-violating
top decays if the decay products contain two leptons plus some jets.

\section{CONCLUSION}
In the minimal supersymmetric standard model the R-parity
violating interactions can induce various exotic three-body decays
for the top quark, which might be accessible at the LHC. We
collectively checked these decays, which include the tree-level
processes $t \to c \ell^-_i \ell^+_j$ ($\ell_i=e, \mu, \tau$) and
$t \to c d_i \bar{d}_j$ ($d_i=d, s, b$), as well as the
loop-induced processes $t \to c g X$ ($X=g, \gamma, Z, h$). We
found that the weakly constrained R-violating couplings can make
the decay branching ratios quite sizable, some of which already
marginally reach the sensitivity of the Tevatron collider and can
be explored at the LHC with better sensitivity.

\section*{Acknowledgement}
This work was supported  by the National Natural
Science Foundation of China (NNSFC) under Nos. 10821504,
10725526 and 10635030.

\appendix
\section{Expressions of loop-induced effective vertices}
Here we list the expressions for the L-violating contributions to the effective vertex $tcg$
in Fig.\ref{fig2}. We also present the results for  the effective vertices $tc\gamma$, $tcZ$
and $tch$, whose Feynman diagrams are similar to  Fig.\ref{fig2}.
Their expressions are given by
\begin{eqnarray}
\Gamma_{\mu}^{tcg}&=& \Gamma_{\mu}^{tcg}(\tilde {l}_i^L)+
                      \Gamma_{\mu}^{tcg}(\tilde {d}_k^R),~~
\Gamma_{\mu}^{tcZ}= \Gamma_{\mu}^{tcZ}(\tilde {l}_i^L)+
                      \Gamma_{\mu}^{tcZ}(\tilde {d}_k^R),\\
\Gamma_{\mu}^{tc\gamma}&=& \Gamma_{\mu}^{tc\gamma}(\tilde {l}_i^L)+
                      \Gamma_{\mu}^{tc\gamma}(\tilde {d}_k^R),~~
\Gamma_{\mu}^{tch}= \Gamma_{\mu}^{tch}(\tilde {l}_i^L)+
                      \Gamma_{\mu}^{tch}(\tilde {d}_k^R).
\end{eqnarray}
where $\tilde {l}_i^L$ and $\tilde {d}_k^R$ denote the L-violating loop contributions
by exchanging respectively sleptons $\tilde {l}_i^L$ and squark $\tilde {d}_k^R$, given by
\small
\begin{eqnarray}
\Gamma_{\mu}^{tcg}(\tilde {l}_i^L)&=& ag_s[C^1_{\alpha\beta}\gamma^\alpha\gamma_\mu\gamma^\beta P_L
    -C^1_\alpha (p_t\!\!\! \!\slash -p_c\!\!\! \!\slash )\gamma_{\mu} \gamma^\alpha P_L
   +\frac{1}{m_t^2}\gamma^\mu p_t\!\!\! \!\slash\gamma^\alpha B^1_{\alpha}P_L\nonumber\\
  &&  -\frac{1}{m_t^2}(\gamma^\alpha p_c\!\!\! \!\slash \gamma_\mu P_L
     + m_t\gamma^\alpha \gamma_\mu P_R)B^2_{\alpha}   ]\\
\Gamma_{\mu}^{tcg}(\tilde {d}_k^R)&=& ag_s[-2C^4_{\alpha\mu}\gamma^\alpha P_L
    +C^4_{\alpha}(p_t+p_c)_\mu\gamma^\alpha P_L
    +\frac{1}{m_t^2}\gamma^\mu p_t\!\!\! \!\slash\gamma^\alpha B^3_{\alpha}P_L\nonumber\\
    && -\frac{1}{m_t^2}(\gamma^\alpha p_c\!\!\! \!\slash \gamma_\mu P_L
     + m_t\gamma^\alpha \gamma_\mu P_R)B^4_{\alpha}   ]\\
\Gamma_{\mu}^{tcZ}(\tilde {l}_i^L)&=& ae\biggl\{\frac{s_W}{3c_W}
   [C^1_{\alpha\beta}\gamma^\alpha\gamma_\mu\gamma^\beta
  -C^1_\alpha (p_t\!\!\! \!\slash -p_c\!\!\! \!\slash )\gamma_{\mu} \gamma^\alpha ]P_L\nonumber\\
&&  +\frac{1}{2s_Wc_W}(1-\frac{4}{3}s_W^2) \frac{1}{m_t^2}\gamma^\mu
     p_t\!\!\! \!\slash\gamma^\alpha B^1_{\alpha}P_L\nonumber\\
&& +(\frac{s_W}{c_W}-\frac{1}{2s_Wc_W})[2C^2_{\alpha\mu}\gamma^\alpha
    -C^2_{\alpha}(p_t+p_c)_\mu\gamma^\alpha ]P_L\nonumber\\
&&  - \frac{1}{2s_Wc_W}\frac{1}{m_t^2}[(1-\frac{4}{3}s_W^2)
       \gamma^\alpha p_c\!\!\! \!\slash \gamma_\mu P_L
      -\frac{4}{3}s_W^2 m_t\gamma^\alpha \gamma_\mu P_R)]B^2_{\alpha}  \biggl\}
\end{eqnarray}
\begin{eqnarray}
\Gamma_{\mu}^{tcZ}(\tilde {d}_k^R)&=& ae\biggl\{\frac{(1-2s_W^2)}{2s_Wc_W}
  [C^3_{\alpha\beta}\gamma^\alpha\gamma_\mu\gamma^\beta
  -C^3_\alpha (p_t\!\!\! \!\slash -p_c\!\!\! \!\slash )\gamma_{\mu} \gamma^\alpha ]P_L\nonumber\\
&& +\frac{s_W}{3c_W}[-2C^4_{\alpha\mu}\gamma^\alpha
    +C^4_{\alpha}(p_t+p_c)_\mu\gamma^\alpha ]P_L
+\frac{1}{2s_Wc_W}(1-\frac{4}{3}s_W^2) \frac{1}{m_t^2}\gamma^\mu
     p_t\!\!\! \!\slash\gamma^\alpha B^3_{\alpha}P_L\nonumber\\
&&  - \frac{1}{2s_Wc_W}\frac{1}{m_t^2}[(1-\frac{4}{3}s_W^2)
       \gamma^\alpha p_c\!\!\! \!\slash \gamma_\mu P_L
      -\frac{4}{3}s_W^2 m_t\gamma^\alpha \gamma_\mu P_R)]B^4_{\alpha}  \biggl\}\\
\Gamma_{\mu}^{tc\gamma}(\tilde {l}_i^L)&=& ae\biggl\{-\frac{1}{3}
[C^1_{\alpha\beta}\gamma^\alpha\gamma_\mu\gamma^\beta
  -C^1_\alpha (p_t\!\!\! \!\slash -p_c\!\!\! \!\slash )\gamma_{\mu} \gamma^\alpha ]P_L
 -[2C^2_{\alpha\mu}\gamma^\alpha
    -C^2_{\alpha}(p_t+p_c)_\mu\gamma^\alpha ]P_L\nonumber\\
&&  +\frac{2}{3}\frac{1}{m_t^2}\gamma^\mu
     p_t\!\!\! \!\slash\gamma^\alpha B^1_{\alpha}P_L
 - \frac{2}{3}\frac{1}{m_t^2}[\gamma^\alpha p_c\!\!\! \!\slash \gamma_\mu P_L
      + m_t\gamma^\alpha \gamma_\mu P_R)]B^2_{\alpha}  \biggl\}\\
\Gamma_{\mu}^{tc\gamma}(\tilde {d}_k^R)&=& ae\biggl\{[C^3_{\alpha\beta}\gamma^\alpha\gamma_\mu\gamma^\beta
  -C^3_\alpha (p_t\!\!\! \!\slash -p_c\!\!\! \!\slash )\gamma_{\mu} \gamma^\alpha ]P_L
 -\frac{1}{3}[-2C^4_{\alpha\mu}\gamma^\alpha
    +C^4_{\alpha}(p_t+p_c)_\mu\gamma^\alpha ]P_L\nonumber\\
&&  +\frac{2}{3}\frac{1}{m_t^2}\gamma^\mu
     p_t\!\!\! \!\slash\gamma^\alpha B^3_{\alpha}P_L
  - \frac{2}{3}\frac{1}{m_t^2}[\gamma^\alpha p_c\!\!\! \!\slash \gamma_\mu P_L
      + m_t\gamma^\alpha \gamma_\mu P_R)]B^4_{\alpha}  \biggl\}
\end{eqnarray}
\begin{eqnarray}
\Gamma_{\mu}^{tch}(\tilde {l}_i^L)&=&
ae\biggl\{-m_{d_k}Y_d[2C^1_\alpha\gamma^\alpha-C^1_0(p_t\!\!\! \!\slash -p_c\!\!\! \!\slash )]P_L
   -\frac{1-\frac{1}{2}s_W^2}{s_Wc_W}m_Z\sin(\alpha+\beta)C^2_\alpha\gamma^\alpha P_L\nonumber\\
&&  -Y_t\frac{1}{m_t^2}[\gamma^\alpha p_c\!\!\! \!\slash P_R
         +m_t\gamma^\alpha P_L]B^2_\alpha \biggl\}\\
\Gamma_{\mu}^{tch}(\tilde {d}_k^R)&=&
ae\biggl\{-m_{l_i}Y_l[2C^3_\alpha\gamma^\alpha-C^3_0(p_t\!\!\! \!\slash -p_c\!\!\! \!\slash )]P_L
   +\frac{s_W}{3c_W}m_Z\sin(\alpha+\beta)C^2_\alpha\gamma^\alpha P_L\nonumber\\
&&  -Y_t\frac{1}{m_t^2}[\gamma^\alpha p_c\!\!\! \!\slash P_R
         +m_t\gamma^\alpha P_L]B^4_\alpha \biggl\}
\end{eqnarray}
\normalsize
with $a=i \lambda^{\prime}_{i3k}\lambda^{\prime}_{i2k}/(16\pi^2)$,
  $s_W=\sin\theta_W$, $c_W=\cos\theta_W$, and $p_t$ and $p_c$ denoting
respectively the momenta of the top and charm quark,
and the Yukawa couplings given by
\begin{eqnarray}
 Y_d= \frac{m_d\sin\alpha}{2m_W\sin\theta_W\cos\beta},~~
    Y_l= \frac{m_l\sin\alpha}{2m_W\sin\theta_W\cos\beta},~~
    Y_t =\frac{m_t\cos\alpha}{2m_W\sin\theta_W\sin\beta}
\end{eqnarray}
For the loop functions B and C in Eqs.(A1-A8), we adopt the
definition in \cite{loop} and use LoopTools \cite{Hahn} in the
calculations. The loop functions' dependence is given by
\begin{eqnarray}
&& C^1=C(-p_t,p_c,m^2_{d_k},m^2_{\tilde{l}_i^L},m^2_{d_k}),
   ~C^2=C(-p_t,p_t-p_c,m^2_{d_k},m^2_{\tilde{l}_i^L},m^2_{\tilde{l}_i^L}),\\
&& C^3=C(-p_t,p_c,m^2_{l_i},m^2_{\tilde{d}_k^R},m^2_{l_i}),
   ~C^4=C(-p_t,p_t-p_c,m^2_{l_i},m^2_{\tilde{d}_k^R},m^2_{\tilde{d}_k^R})~,\\
&& B^1=B(-p_t,m^2_{d_k}, m^2_{\tilde{l}_i^L}),
   ~B^2=B(-p_c,m^2_{d_k}, m^2_{\tilde{l}_i^L})~,\\
&& B^3=B(-p_t,m^2_{l_i}, m^2_{\tilde{d}_k^R}),
   ~B^4=B(-p_c,m^2_{l_i}, m^2_{\tilde{d}_k^R})~.
\end{eqnarray}


\begin{thebibliography}{99}
\bibitem{topreview} For top quark reviews, see, e.g.,
               W. Bernreuther, \JPG 35, 083001,(2008)
               D. Chakraborty, J. Konigsberg, D. Rainwater,
                         Ann. Rev. Nucl. Part. Sci. {\bf 53}, 301  (2003);
               E.~H.~Simmons, hep-ph/0211335; hep-ph/0011244;
               C.-P. Yuan,  hep-ph/0203088;
               S. Willenbrock, hep-ph/0211067;
               M. Beneke {\it et al.}, hep-ph/0003033;
               T. Han, arXiv:0804.3178;
               For model-independent new physics in top quark, see, e.g.,
               C. T. Hill, S. J. Parke, \PRD49, 4454 (1994);
               K. Whisnant {\it et al.},  \PRD56, 467 (1997);
               J. M. Yang, B.-L. Young, \PRD56, 5907 (1997);
               K. Hikasa {\it et al.}, \PRD58, 114003 (1998);
               J. A. Aguilar-Saavedra, arXiv:0811.3842.
\bibitem{tcvh-sm} For top FCNC in the SM, see,
                  G.~Eilam, J.~L.~Hewett, A.~Soni, \PRD44, 1473 (1991);
                   B.~Mele, S.~Petrarca,  A.~Soddu, \PLB435, 401 (1998);
                   A.~Cordero-Cid {\it et al.}, \PRD73, 094005 (2006);
                  G.~Eilam, M.~Frank, I.~Turan, \PRD73, 053011 (2006).
\bibitem{tcv-mssm-1} For top FCNC in R-conserving MSSM, see, e.g.,
    C.~S.~Li, R.~J.~Oakes, J.~M.~Yang, \PRD49, 293 (1994);
    G.~Couture, C.~Hamzaoui, H.~Konig, \PRD52, 1713 (1995);
    J.~L.~Lopez, D.~V.~Nanopoulos, R.~Rangarajan, \PRD56, 3100  (1997);
    G.~M.~de Divitiis, R.~Petronzio, L.~Silvestrini, \NPB504, 45 (1997);
    C.~S.~Li, L.~L.~Yang, L.~G.~Jin, \PLB599, 92 (2004);
    M.~Frank, I.~Turan, \PRD74, 073014 (2006);
    J.~M.~Yang, C.~S.~Li, \PRD49, 3412 (1994);
    J.~Guasch, J.~Sola, \NPB562, 3 (1999);
    J. Guasch, {\it et al.}, hep-ph/0601218;
    J. M. Yang, Annals Phys. {\bf 316}, 529 (2005);
                Int. J. Mod. Phys. A23, 3343 (2008);
    J.~Cao, {\it et al.}, \NPB651, 87 (2003); \PRD74, 031701 (2006);
                          \PRD75, 075021 (2007).
\bibitem{tcv-mssm-2} For top FCNC in R-violating MSSM, see,
    J.~M.~Yang, B.-L.~Young, X.~Zhang, \PRD58, 055001 (1998);
    G. Eilam, {\it et al.}, \PLB510, 227 (2001);
    J. Cao, {\it et al.},  \PRD79, 054003 (2009);
\bibitem{tcv-tc2}  For top FCNC in TC2, see, e.g.,
   H. J. He and C. P. Yuan, \PRL83, 28(1999);
   G. Burdman, \PRL83,2888(1999);
   X.~L. Wang {\it et al.}, \PRD50, 5781 (1994);
   C.~Yue, {\it et al.}, \PLB 496, 93 (2000);
   J. Cao, {\it et al.}, \PRD67, 071701 (2003); \PRD70, 114035 (2004);
                         \EPJC41, 381 (2005); \PRD76, 014004 (2007);
   H. J. Zhang, \PRD77, 057501 (2008);
   G. L. Liu, H. J. Zhang, Chin. Phys. C 32, 597 (2008) [arXiv:0708.1553];
   G. L. Liu, arXiv:0903.2619.
\bibitem{hosch} M. Hosch, {\it et al.}, \PRD58, 034002 (1998);
                G. Mahlon, G. L. Kane, \PRD55, 2779 (1997);
                S. Mrenna, C.P. Yuan, \PLB367, 188 (1996);
                J. Sender, \PRD54, 3271 (1996).
\bibitem{rv-t-prod} A. Datta,  {\it et al.}, \PRD56, 3107 (1997);
                     R. J. Oakes, {\it et al.}, \PRD57, 534 (1998);
                     P. Chiappetta, {\it et al.}, \PRD61, 115008 (2000);
                      D.~K. Ghosh, S.~Raychaudhuri, K.~Sridhar, \PLB396, 177 (1997);
                 K. Hikasa, J. M. Yang, B.-L. Young, \PRD60, 114041 (1999);
                 P. Li, {\it et al.}, \EPJC51, 163 (2007).
\bibitem{t-3decay-mssm1} J. Guasch, J. Sola, Z. Phys. C74, 337 (1997);
                         for model-independent study, see,
                         J. Drobnak, S. Fajfer, J. F. Kamenik, arXiv:0812.0294 [hep-ph].
\bibitem{t-3decay-mssm2} A. Belyaev, J. R. Ellis, S. Lola, \PLB484, 79 (2000).
\bibitem{rv-t-decay} K.J. Abraham, {\it et al.}, \PRD63, 034011 (2001); \PLB514, 72 (2001).
\bibitem{t-3decay-2hdm}  C.S. Li, {\it et al.}, \PRD51, 4971 (1995).
\bibitem{Rpv}    For example, see,
                 C. S. Aulah, R. N. Mohapatra, \PLB119, 316 (1982);
                 L. J. Hall, M. Suzuki, \NPB231, 419 (1984);
                  S. Dawson, \NPB261, 297, (1985);
                 R. Barbieri, A. Masiero, \NPB267, 679 (1986);
                 S. Dimopoulos, L. J. Hall, \PLB196, 135 (1987);
                 V.~Barger, G.~F.~Giudice, T.~Han, \PRD40, 2987 (1989);
\bibitem{proton}  See, e.g.,
                  C.~Carlson, P.~Roy and M.~Sher, \PLB357, 99 (1995);
                 A.~Y. Smirnov and F.~Vissani, \PLB380, 317 (1996).
\bibitem{bound}  M. Chemtob, Prog. Part. Nucl. Phys. {\bf 54}, 71 (2005) [hep-ph/0406029];
                 R. Barbier {\it et al.}, Phys. Rept. 420, 1 (2005).
\bibitem{CDF-mt}  Y.-C. Chen, for the CDF and D0 Collaborations, arXiv:0805.2350 [hep-ex].
\bibitem{pdg}  C. Amsler {\it et al.}, Particle Data Group, \PLB667, 1 (2008).
\bibitem{CDF}  T. Aaltonen {\it et al.}, CDF Collaboration, arXiv: 0811.2512.
\bibitem{LEP}  P.~Achard {\it et al.}, L3 Collaboration, \PLB580, 37 (2004).
\bibitem{CDF-tt} A. Lister, for the CDF and D0 Collaborations, arXiv:0810.3350 [hep-ex].
\bibitem{QCD-tt}    N. Kidonakis and R. Vogt, \PRD78, 074005 (2008).
\bibitem{tt-susy}  C.S. Li, {\it et al.}, \PRD52, 5014 (1995); \PLB379, 135 (1996);
                                          \PRD52, 1541 (1995); \PRD54, 4380 (1996)
    J. Kim,  {\it et al.}, \PRD54, 4364 (1996);
    S. Alam, K. Hagiwara and S. Matsumoto, \PRD55, 1307 (1997);
    Z. Sullivan, \PRD56, 451 (1997);
    W. Hollik, W.M. Mosle and D. Wackeroth, \NPB516, 29 (1998).
\bibitem{loop}  B.~A.~Kniehl, Phys. Rept. 240, 211 (1994).
\bibitem{Hahn} T.~Hahn, M.~Perez-Victoria, Comput.\ Phys.\ Commun.\ {\bf 118}, 153 (1999);
               T.~Hahn,  Nucl.\ Phys.\ Proc.\ Suppl.\ {\bf 135}, 333 (2004).
\bibitem{rv-t-decay-lhc} A. Belyaev, M-H. Genest, C. Leroy, R. Mehdiyev,
                         JHEP 0409, 012 (2004).
\end{thebibliography}
\end{document}